\documentclass[a4paper,11pt,onecolumn]{article}
\usepackage[dvips]{graphicx}
\usepackage{latexsym}
\title{Wealth distribution in a System with Wealth-limited Interactions}
\author{Marisciel L. Palima and \ Eduardo J. David} 
\date{Submitted: October 4, 2007}

\begin{document}
\maketitle

\begin{abstract}
We model a closed economic system with interactions that generates the features of empirical wealth distribution across all wealth brackets, namely a Gibbsian  trend in the lower and middle wealth range and a Pareto trend in the higher range, by simply limiting the an agents' interaction to only agents with nearly the same wealth.     To do this, we introduce a parameter $\beta$  that limits the range on the wealth of a partner with which an agent is allowed to interact.  We show that this wealth-limited interaction is enough to distribute wealth in a purely power law trend.  If the interaction is not wealth limited, the wealth  distribution is expectedly Gibbsian.  The value of $\beta$   where the transition from a purely Gibbsian law to a purely power law distribution  happens depends on whether the choice of  interaction partner is mutual nor not.  For a  non-mutual choice, where the richer agent gets to decide, the transition happens at  $\beta  = 1.0$.  For a mutual choice, the transition is at  $\beta = 0.60$.  In order to generate a mixed Gibbs-Pareto distribution, we apply another wealth-based rule that depends on the parameter $w_{limit}$.  An agent whose wealth is below $w_{limit}$ can choose any partner to interact with, while an agent whose wealth is above $w_{limit}$ is subject to the wealth-limited range in his choice of partner.  A Gibbs-Pareto distribution appears if both these wealth-based rules are applied.
\end{abstract}
\section{Introduction}
\label{intro}
We are interested in the steady-state income distribution for a closed economy. Several studies have dealt with this question using a group of interacting agents with different parameters like saving propensity, altruism, and preferential attachments \cite{Ref1}-\cite{Ref4}. For simple interactions, the wealth distribution is shown to be an exponential or a Gibbs distribution \cite{Ref2,Ref3,Ref8}.

Empirical data shows otherwise.  As early as the 1890's, Pareto has shown that the cumulative income distribution has a power law trend, $P(w) \propto w^{-\gamma}$. The exponent $\gamma$, also named as the Pareto index or Pareto exponent, for several countries was found to be $1.2 \leq \gamma \leq 1.9$ \cite{Ref2,Ref8}. 
However, recent studies show that the Pareto tendency is correct only for the higher income brackets.  Real data for the middle and lower brackets shows an exponential trend \cite{Ref5}-\cite{Ref7}. In this work, we show that we can simulate a Gibbsian distribution as well as a Pareto distribution with a suitable change in the rule guiding the transaction.  We also show how a mixed exponential and Pareto law, which is descriptive of empirical data, can also be generated.
	Previous works have also produced a distribution that is comparable to empirical data. In the work of Scafetta et al. \cite{Ref11}, the distribution was produced using three parameters (social index, poverty index, and investment index) in two mechanisms (trade and investments). In the recent work by Scafetta \cite{Ref12}, the Gibbs-Pareto distribution was obtained by assuming that each agent has its random tendency to risk a certain portion of one's wealth in every transaction.  This tendency is called saving propensity.  In this work, we show that there is another mechanism en route to a mixed Gibbs-Pareto distribution that is based on the assumption that agents' restrict their interactions based on wealth comparisons.  Moreover, we emphasize in this work that although wealth-limited interactions happen in actual economic scenarios, corresponding rules are not commonly included in agent-based models of wealth distribution.
We found out that the single most important rule that gives an anomalous distribution is changing the partnering criterion. By simply limiting the agents' interaction to other agents of nearly the same wealth, a power law distribution maybe obtained. And then, by simply mixing the rules, an exponential shoulder in the income distribution is established.

\section{Methods}
\label{method}
Our model consists of $10^4$ interacting agents characterized by wealth $w$ and transaction range $\beta$. Unlike in the work done by Laguna et. al. \cite{Ref8}, we assume that the notion of an "economic class", represented by the parameter $\beta$, is relative, meaning it is dependent on the amount of wealth the agent currently has. Each agent initially receives a fixed amount $w = 1000$. We only simulated a closed economy since we observed that using random parameters for an open economy, a system where each agent receives an additional wealth after some time steps, produces the same result as the closed economy, which is an exponential wealth distribution. At each time step an agent $i$ is chosen randomly and also a second agent, $j$, is picked at random within the whole population or at random within an agent's economic class. For each agent, the value $[w \pm \beta w]$ represents the transaction range.
Wealth exchange is done by first choosing a random value for the exchange money, $dw = ran[0, w]$, for each agent. This ensures that no agent would have a negative money, or debt, and to ensure that the system remains conservative.  After the exchange, the wealth of the agents would be equal to

\begin{eqnarray}
w_w=w_w+dw_l \\
w_l=w_l-dw_l 
\nonumber
\end{eqnarray}

where the subscripts $w$ and $l$ denotes the winner and the loser respectively. The winner is determined at random without favoring any agent \cite{Ref8}-\cite{Ref10} and receives the amount $dw_l$.  Also, each agent is only allowed to interact at most once per iteration. Agents with $0$ wealth were still included in the exchange. Analysis of the stationary state was done after $500$ iterations.

The simplest model, where the selection of partners is random, was done by shuffling the entire population and then picking agent($i$) as the first agent and agent($i$ + 1) as the second, where  $i = 1, 3, 5,\dots,   {N - 1}$. In this method, all agents participate in the wealth exchange process. We have also studied an open economy using this rule. In this system, we add a fixed amount of wealth to every agent after a fixed amount of time.

The economic class rule was implemented by including a range which determines which of the agents i can choose to be the partner. This range is dependent on the agent $i$'s current wealth, i.e. $w_j$ must be within $[w_i \pm \beta w_i]$. The values of $\beta$ used ranges from $0.1$ to $1.0$ in increments of $0.1$. This model represents the case when the richer of the agents choose his/her trading partner.

We consider the case of a more restrictive economic class rule wherein both agents must fall within each other's range. In this new rule, agent $i$ is determined randomly and then agent $j$ would be chosen using the same rule above. If $w_j$ falls within $[w_i \pm \beta w_i]$, $w_i$ would then be tested against agent $j'$s range. If $w_i$ falls within the range then an agent $i$ and $j$ are paired, if not, a new agent $i$ would be chosen again randomly from the agents which did not yet participate in the trading for that iteration. This process would be repeated until all the agents have already interacted or until all remaining agent cannot find an appropriate partner. The value of $\beta$ was also tested from $0.1$ to $1.0$.

A mixing of the random rule and the economic rule was also simulated. For this rule, we assigned a fixed value $w_{limit}$ which represents the separation between the poor/middle class and the rich. If $w_i < w_{limit}$, then agent $i$ would choose a partner using the random rule and if $w_i > w_{limit}$, agent $i$ would choose following the second, or mutual, economic class rule. These limits represent the case where the rich would only want to interact with other rich agents while the poor and middle class agents don't care about the amount of money their partner has.

\section{Results and Discussion}

% For one-column wide figures 
\begin{figure}
\resizebox{\textwidth}{!}{
\includegraphics{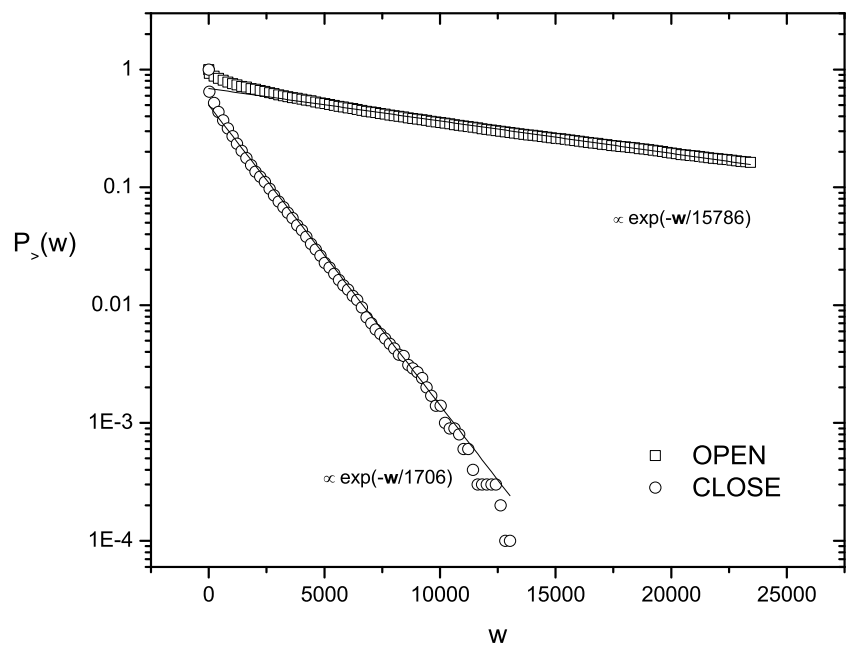}
}
\caption{The semilog plot of wealth distribution for an unrestricted or random choice of interaction partners suggests a Gibbsian trend.}
\label{fig:1}
\end{figure}

	In all distributions, we consider a Pearson correlation factor $R^2>0.95$ to be a good fit. The obtained probability distribution for the random model was  exponential as expected. The distribution plot for an open economy follows the same trend as the closed economy. The plots only differ in the slope, which is expected, since our open economy model can increase the total wealth to a very large amount. The open economy system that we simulated includes increasing the wealth of every agent by $100$ for every $5$ iterations. We also tried increasing the wealth after $10$ iterations and also changing the wealth increase by $200$. The comparison of results for the open and closed economy is shown in Figure \ref{fig:1}.

\begin{figure}
\resizebox{\textwidth}{!}{
\includegraphics{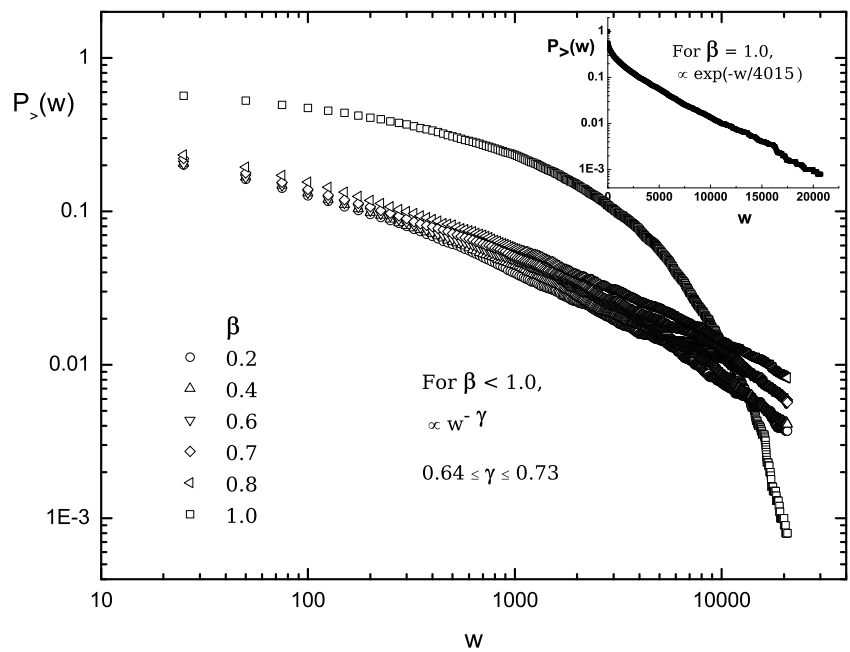}
}
\caption{Log-log plot of wealth distribution at different values of $\beta$ for the non-mutual choice of partners.  For this case, it is up to the richer agent to choose the partner. For $\beta < 1.0 $, the distribution is Pareto with the exponent $\gamma$ within $[0.64,0.73]$.}
\label{fig:2}
\end{figure}

	The results when we let the first agent choose a partner within his economic class are shown in Figure \ref{fig:2}. Here we found a limit where the distribution shifts to a pure power law. Above $\beta = 1.00$, the distribution is exponential. For values of $\beta < 1.00$, the data can is best fitted with a power law function. When $\beta < 1.0$, the distribution shifts to a power law behavior. At low values of $\beta$, the poor agents were left to trade with other poor agents or at most with middle class agents. A consequence of this is the difficulty for the poor to improve their state which produces a high inequality between the lower and upper classes especially at very low values of $\beta$. 

% For one-column wide figures 

\begin{figure}
\resizebox{\textwidth}{!}{
\includegraphics{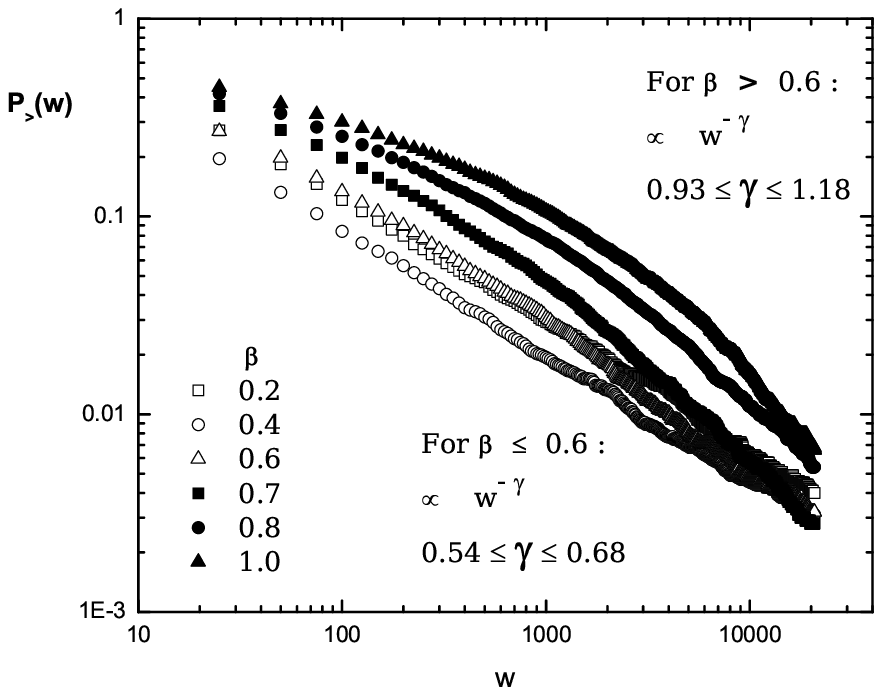}
}
\caption{Log-log plot of wealth distribution at different values of $\beta$ for the mutual choice of partners.  For this case, both partners must be within each other's range. The plots are best-fitted with a power law function, with the exponent $\gamma$ within $[0.54,0.68]$ for $\beta \leq 0.6$ and within $[0.93,1.18]$ for $\beta < 0.6$.  A phase transition happens for $\beta=0.6$.}
\label{fig:3}
\end{figure}

	Figure \ref{fig:3} gives the plot for the case where both agents must be within each others range before they interact. For this case, we observed the mininum value of $\beta$ for the distribution to shift to a power law is high compared to the previous case. The maximum value obtained in this case was $0.6 < \beta < 0.7$. The distribution has a power law behavior for values of $\beta < 0.6$. The value needed for $\beta$ to shift the distribution increases because unlike in the first rule, where the poor agents still have a chance to obtain a portion of the rich agents wealth, here poor agents were completely left to interact with other poor agents. This further increases the inequality between the agents. 

% For one-column wide figures 

% For one-column wide figures 
\begin{figure}
\resizebox{\textwidth}{!}{
\includegraphics{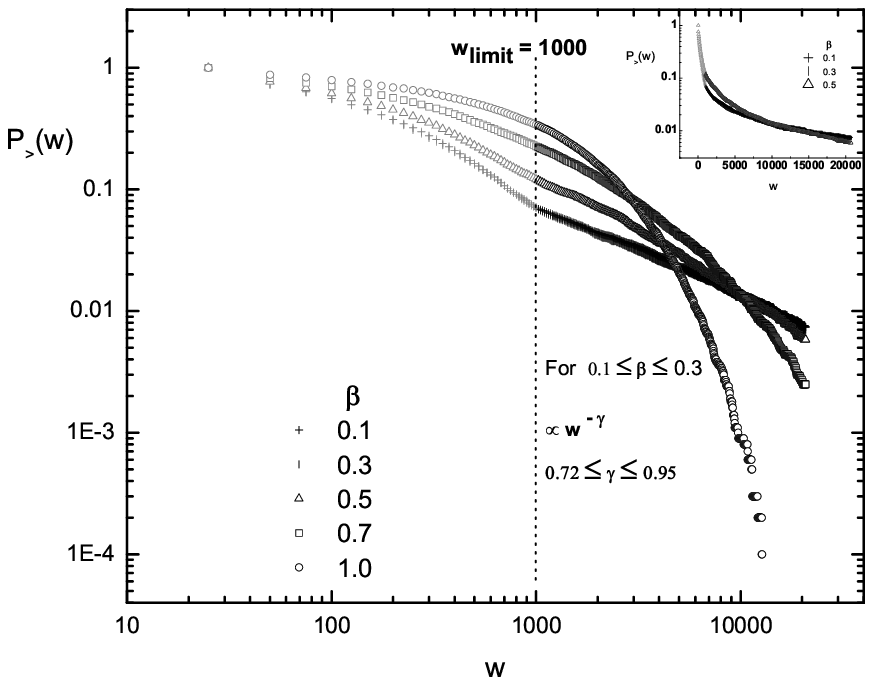}
}
\caption{Semi-log plot of wealth distribution at different values of $\beta$ for the mixed random and $\beta$-based choice of partners. The separation between the exponential and power law behavior occurs near the specified $w_{limit}$=1000.}
\label{fig:4}
\end{figure}

	Figure \ref{fig:4} shows the distribution function when we mixed the random rule with the mutual economic range rule. When $\beta > 0.60$, the distribution can be approximated to be purely exponential. However, at values of $\beta < 0.60$, the distribution can be separated into two regimes. At values lower than the $w_{limit}$, the distribution have a dominant exponential dependence while at values higher than this limit, a power law behavior is more dominant. The obtained Pareto index using these rules is $0.6 < \gamma < 0.8$, which is significantly different from empirical values where the Pareto exponent $\gamma \sim 2$.  This suggest that there are other mechanisms that must be considered to obtain a steeper power law behavior in order to better fit the real data.  For example, as Chatterjee reported \cite{Ref2} a wealth distribution is obtained with the proper Pareto exponent when the saving propensity for each agent is not zero.  Nevertheless, we consider our suggested wealth-limited interaction as a more basic mechanism en route to a Pareto wealth distribution than the saving propensity assumption.  Moreover, the wealth-based rule is more realistic rule and is easier to implement than the rule based on saving propensity, which is harder to determine in actual situations.

\section{Conclusion}
\label{conc}
	In this work, the wealth distribution in a closed economy was studied using different rules for pairing the agents. By changing only this rule, we were able to generate an exponential or Gibbs distribution, a Pareto distribution, and also the Gibbs-Pareto distribution. The Gibbs wealth distribution can be shifted to a Pareto distribution by imposing that an agent can only interact with other agents who belong to his economic class. Using the economic class rule, where the richer of the agents have a privilege to choose his partner, the obtained maximum transaction range was $\beta = 1.0$ which corresponds to ${} \pm 10\%$ of the agents current wealth. However, if we let the agents fall within each others range before they can interact, the maximum value of $\beta$ before the distribution shifts to a power law decreases to $0.60$.

By imposing that the poor and middle class agents have different behavior when choosing a partner than that of the rich agents, an exponential or Gibbs distribution is obtained at wealth $w< w_{limit}$, and a power law distribution at wealth $w> w_{limit}$. However, the power law exponent, or Pareto index, obtained here was $0.6 < \gamma < 0.8$. In order to give a Pareto index closer to the empirical data, other mechanisms must be considered like considering interactions subject to the saving propensity $\lambda$. 
  
We showed that there is another route to a Pareto wealth distribution via a wealth-limited interation.  This is an arguably a more realistic mechanism compared to the saving propensity rule because in actual conditions, it is easier to measure and compare the agents' wealth than to measure an agent's saving propensity. Limiting an economic transaction to agents of roughly the same capacity is similar to categorizing financial derivatives as low, medium and high risks and enticing clients to buy one of such derivatives according to their economic means.  In the same fashion, imposing the $w_{limit}$, is like setting a minimum joining fee before a customer can participate in a high stakes transaction.  Our results show that these rules on choosing a partner based on their current wealth is an important factor that shapes the wealth distribution.  As far as we have searched, these simple rules have not been considered in previous agent-based models of wealth distribution.

% Non-BibTeX users please use

\end{document}